\documentclass[10pt,twocolumn,oneside,a4paper,final]{IEEEtran}
\usepackage[utf8]{inputenc}
\def\BibTeX{{\rm B\kern-.05em{\sc i\kern-.025em b}\kern-.08em
    T\kern-.1667em\lower.7ex\hbox{E}\kern-.125emX}}

\usepackage{graphicx}
\usepackage{color}
\usepackage{cite}
\usepackage{amsmath}
\usepackage[caption=false]{subfig}
\usepackage{amssymb}
\usepackage{comment}
\usepackage{mathtools}
\usepackage{tabulary}
\usepackage{acronym}
\usepackage{upgreek}
\usepackage{algorithm}
\usepackage{algpseudocode}
\usepackage{placeins}
\usepackage[super]{nth}
\usepackage{dsfont}
\usepackage{hyperref}

\DeclareMathOperator{\relu}{ReLU}
\DeclareMathOperator*{\argmax}{arg\!\max}

\newcommand{\myvec}[1]{\boldsymbol{#1}}

\newacro{ANN}[ANN]{Artificial Neural Network}
\acrodefplural{ANN}[ANNs]{Artificial Neural Networks}

\newacro{RIS}[RIS]{Reconfigurable Intelligent Surface}
\acrodefplural{RIS}[RISs]{Reconfigurable Intelligent Surfaces}

\newacro{6G}[6G]{Sixth Generation}
\newacro{SNR}[SNR]{Signal-to-Noise Ratio}
\newacro{SINR}[SINR]{Signal-to-Interference-plus-Noise Ratio}
\newacro{TX}[TX]{Transmitter}

\newacro{RX}[RX]{Receiver}
\acrodefplural{RX}[RXs]{Receivers} 

\newacro{TVC}[TVC]{Time Varying Channel}
\acrodefplural{TVC}[TVCs]{Time Varying Channels}

\newacro{BS}[BS]{Base Station}
\newacro{UE}[UE]{User Equipment}
\acrodefplural{UE}[UEs]{Users Equipments}

\newacro{MSE}[MSE]{Mean Squared Error}

\newacro{MAC}[MAC]{Medium Access Control}

\newacro{NMSE}[NMSE]{Normalized Mean Squared Error}
\newacro{EM}[EM]{Electro-Magnetic}
\newacro{CSI}[CSI]{Channel State Information}
\newacro{MISO}[MISO]{Multiple-Input Single-Output}
\newacro{SISO}[MISO]{Single-Input Single-Output}
\newacro{NOMA}[NOMA]{Non-Orthogonal Multiple Access}
\newacro{OFDM}[OFDM]{Orthogonal Frequency-Division Multiplexing}
\newacro{QoS}[QoS]{Quality of Service}

\newacro{UAV}[UAV]{Unmanned Autonomous Vehicle}
\acrodefplural{UAV}[UAVs]{Unmanned Autonomous Vehicles}

\newacro{IoT}[IoT]{Internet of Things}
\newacro{LOS}[LOS]{Line Of Sight}
\newacro{NLOS}[NLOS]{Non-Line Of Sight}

\newacro{AoA}[AoA]{Angle of Arrival}
\acrodefplural{AoA}[AoAs]{Angles of Arrival}

\newacro{AoD}[AoD]{Angle of Departure}
\acrodefplural{AoD}[AoDs]{Angles of Departure}

\newacro{AWGN}[AWGN]{Additive White Gaussian Noise}
\newacro{SL}[SL]{Supervised Learning}
\newacro{RL}[RL]{Reinforcement Learning}
\newacro{DRL}[DRL]{Deep Reinforcement Learning}

\newacro{MDP}[MDP]{Markov Decision Process}
\acrodefplural{MDP}[MDPs]{Markov Decision Processes} 

\newacro{POMDP}[POMDP]{Partially Observable Markov Decision Process}
\acrodefplural{POMDP}[POMDPs]{Partially Observable Markov Decision Processes}

\newacro{DQN}[DQN]{Deep Q-Network}
\newacro{D$^3$QN}[D$^3$QN]{Duelling Double Deep Q Network}
\newacro{PG}[PG]{Policy Gradient}
\newacro{PPO}[PPO]{Proximal Policy Optimization}
\newacro{DDPG}[DDPG]{Deep Deterministic Policy Gradient}
\newacro{UCB}[UCB]{Upper Confidence Bound}
\newacro{GPU}[GPU]{Graphical Processing Unit}
\newacro{PDS}[PDS]{Post Decision State}
\newacro{TD}[TD]{Temporal Difference}
\newacro{MAML}[MAML]{Multi-Agent Reinforcement Learning}
\newacro{LSTM}[LSTM]{Long Short-Term Memory}
\newacro{EE}[EE]{Energy Efficiency}
\newacro{SE}[SE]{Spectral Efficiency}
\newacro{MIMO}[MIMO]{Multiple-Input Multiple-Output}

\newacro{IID}[IID]{Independent and Identically Distributed}

\newacro{MAB}[MAB]{Multi-Armed Bandits}
\newacro{CB}[CB]{Contextual Bandits}

\newacro{DRP}[DRP]{Deep Reward Prediction}

\newacro{DFT}[DFT]{Discrete Fourier Transform}
\newacro{ML}[ML]{Machine Learning}

\newacro{SGD}[SGD]{Stochastic Gradient Descent}

\newacro{AI}[AI]{Artificial Intelligence}

\title{Online RIS Configuration Learning for Arbitrary Large Numbers of $1$-Bit Phase Resolution Elements}
\author{\IEEEauthorblockN{Kyriakos Stylianopoulos and George C. Alexandropoulos}\\
\IEEEauthorblockA{Department of Informatics and Telecommunications,
National and Kapodistrian University of Athens\\
Panepistimiopolis Ilissia, 15784 Athens, Greece\\
emails: \{kstylianop, alexandg\}@di.uoa.gr
}
\thanks{This work has been supported by the EU H2020 RISE-6G project under grant number 10101701.}\vspace{-0.7cm}}
\begin{document}

\maketitle

\begin{abstract}
Reinforcement Learning (RL) approaches are lately deployed for orchestrating wireless communications empowered by Reconfigurable Intelligent Surfaces (RISs), leveraging their online optimization capabilities. Most commonly, in RL-based formulations for realistic RISs with low resolution phase-tunable elements, each configuration is modeled as a distinct reflection action, resulting to inefficient exploration due to the exponential nature of the search space. In this paper, we consider RISs with $1$-bit phase resolution elements, and model the action of each of them as a binary vector including the feasible reflection coefficients. We then introduce two variations of the well-established Deep Q-Network (DQN) and Deep Deterministic Policy Gradient (DDPG) agents, aiming for effective exploration of the binary action spaces. For the case of DQN, we make use of an efficient approximation of the Q-function, whereas a discretization post-processing step is applied to the output of DDPG. Our simulation results showcase that the proposed techniques greatly outperform the baseline in terms of the rate maximization objective, when large-scale RISs are considered. In addition, when dealing with moderate scale RIS sizes, where the conventional DQN based on configuration-based action spaces is feasible, the performance of the latter technique is similar to the proposed learning approach.
%while at the same time attain performance equal to the original DQN algorithm in small-scale RISs, in which comparisons are tractable.
\end{abstract}
\vspace{-0.2cm}

\vspace{-0.1cm}
\section{Introduction}\label{sec:introduction}
%\vspace{-0.1cm}
The technology of \acp{RIS} has been acknowledged as one of the key ingredients of next $6$-th Generation (6G) of wireless networks~\cite{rise6g}.
Those surfaces consist of potentially large numbers of nearly passive (i.e., without any power amplification) meta-material elements that induce a phase shift to the propagating wireless signals, depending on their internal generalized reflection states \cite{RIS_Overview}. The overall RIS configuration can be intelligently controlled, therefore empowering the wireless environment with dynamic reconfiguration abilities that offer unprecedented benefits in terms of performance indicators and provided services~\cite{RISE6G_COMMAG}.
\par
To fully realize the potential of RISs \cite{WavePropTCCN}, however, their phase configurations need to be carefully selected to serve the underlying system objective. As a result, the problem of phase tuning has been extensively studied using either conventional optimization schemes (e.g., \cite{Peng2021RIS_Opt, Cheng2021RIS_Opt}), or techniques stemming from \ac{ML} \cite{ASBM20, AIRIS}. A distinct sub-domain of the latter, called \ac{DRL} is especially designed for solving online decision problems using learning algorithms that are trained through continuous interactions within a controllable environment.
\ac{DRL} methods targeting \ac{RIS} control cover a great variety of design objectives, including energy efficiency \cite{Lee2020DRL_EE}, resource scheduling \cite{gao2021ResourceAllocation, alhilo2021reconfigurable}, and secrecy rate \cite{Yang2021DRL_for_Secure}.
However, the principal utilization of such \ac{AI}-based orchestrators is for increased spectral efficiency through combinations of analog (\ac{RIS}) and analog/digital (transmit/receive) beamforming \cite{Abdelrahman2020TowardsStandaloneOperation, Huang2020DRL_RIS, Feng2020DRL_MISO, Stylianopoulos2022DeepCB, Huang2021MultiHop, kim2021multiirsassisted }, alongside other considerations such as power allocation \cite{Liu2021MNOMA_deployment} or \ac{UAV} control \cite{Samir2021AgeOfInformation}.
\par
In this paper, we are concerned with the problem of configuring \acp{RIS} that are consisted of large numbers of individually controllable unit-elements through \ac{DRL} orchestration. Future wireless environments are envisioned to deploy multiple operating metasurfaces~\cite{RISE6G_COMMAG}, each one comprised of hundreds or thousands of phase-tunable elements. At the same time, current \ac{RIS} prototypes are predominantly designed with $1$-bit quantized phase shifts per element \cite{RIS_Overview,alexandg_2021}, leading to (base-2) exponential numbers of available RIS configurations. In general, \ac{DRL} agents that are tasked with selecting the discrete phase shifts, treat each of the possible configurations as an individual action (e.g., \cite{Stylianopoulos2022DeepCB,Huang2021MultiHop}) and their training process involves receiving feedback on the selected profile at every iteration. As a result, such algorithms are prone to inefficient search-space-exploration and slow convergence rate due the rapid increase of the cardinality of the action space. Motivated by the fact that for $1$-bit phase-quantized \acp{RIS}, each element's action can be represented by a binary vector, so that each vector element denotes the selection of one of the two available phase shifts, we devised two modified versions of the celebrated \ac{DRL} algorithms \ac{DQN} and \ac{DDPG}, which leverage the binary decomposition of the \ac{RIS} configurations, resulting in both cases to an action space that is linear to the number of RIS elements. This is amenable to tuning each element individually, which leads to a more efficient propagation of feedback information, compared to treating each available overall RIS configuration as an individual action.
\par
The rest of the paper is structured as follows:
Section~\ref{sec:System_Model} describes the considered system setup and operation objective, whereas Section~\ref{sec:methodology} includes the paper's \ac{DRL} formulation and presents the proposed modified versions of the \ac{DQN} and \ac{DDPG} algorithms. These algorithms are numerically evaluated in Section~\ref{sec:Numerical_Evaluation}, while Section~\ref{sec:Discussion} contains a discussion on the proposed methodology, followed by the paper's conclusion in Section~\ref{sec:Conclusion}.
\par
\textbf{Notation:} Bold-faced small and capital letters denote vectors and matrices, respectively, while calligraphy typeface denotes sets.
$[\myvec{x}]_{i}$ denotes the $i$-th element of $\myvec{x}$.
The cardinality of a set $\mathcal{S}$ is expressed as ${\rm card}(\mathcal{S})$,
the ${\rm vec}(\cdot)$ operator vectorizes a matrix in row format,
and ${\rm diag}(\myvec{x})$, for an $n$-dimensional vector $\myvec{x}$, creates an $n \times n$ matrix with the elements of $\myvec{x}$ placed along the main diagonal.
The expectation operation is expressed as $\mathbb{E}\{ \cdot \}$
and ${\rm Real}\{\cdot\}$ (${\rm Imag}\{\cdot\}$) returns the real (imaginary) part of a complex quantity.
Finally, $\jmath\triangleq \sqrt{-1}$.
\vspace{-0.1cm}
\section{System Model and Design Objective}\label{sec:System_Model}
%\vspace{-0.1cm}
In this section, we give a description of the system model that will be considered during the presentation of the \ac{DRL} methods and the numerical evaluation.
Since the aim of this paper is to examine the performance of \ac{ML}-based controllers with large-scale RISs, we have purposely selected a simple system architecture for clarity. More complex system architectures will be studied in the journal version of this work. We assume a \ac{MISO} downlink communication wireless environment that includes a \ac{BS} equipped with $K$ antennas and a single-antenna \ac{UE} which remains at a fixed location. The presence of a blocker is assumed to obstruct the direct link between them. Instead, the communication is enabled by the positioning of an \ac{RIS} consisted of $N$ controllable phase shifting elements. As is typical in the industry \cite{RIS_Overview,alexandg_2021}, we consider a metasurface structure, in which the state $\varphi_i$ of each unit meta-element $i$ $(1 \leq i \leq N)$ can be set to one out of two predefined phases, say $\vartheta_1$ and $\vartheta_2$.
The configuration space of the RIS can thus be defined as $\mathcal{F} = \{\vartheta_1,  \vartheta_2\}^N$ with ${\rm card}(\mathcal{F}) = 2^N$.
Let $\myvec{\phi} \triangleq [ \exp{(j \pi \varphi_1)}, \dots, \exp{(j \pi \varphi_N)} ]^T$ denote the configuration vector of the \ac{RIS} and let $\myvec{\Phi} \triangleq {\rm diag}(\myvec{\phi}) \in \mathbb{C}^{N \times N}$.
By further denoting with $\mathbf{H} \in \mathbb{C}^{K \times N_{\rm T}}$ and $\mathbf{g} \in \mathbb{C}^{1 \times N}$ the channel gain matrices for the \ac{BS}-\ac{RIS} and \ac{RIS}-\ac{UE} wireless links, respectively, the baseband received signal at the \ac{UE} can be expressed as
\begin{equation}
    y = \mathbf{g} \myvec{\Phi} \mathbf{H} \mathbf{v} x + \tilde{n},
\end{equation}
where $\tilde{n}$ models the \ac{AWGN} with zero mean and variance $\sigma^2$, $x$ is the symbol transmitted with power $P$, and $\mathbf{v} \in \mathbb{C}^{N_{\rm T}\times1}$ represents the BS precoding vector.
To focus specifically on the RIS phase configuration control in this paper, we do not consider the design and selection of the precoder as part of the problem formulation, even though it constitutes an important aspect of wireless systems with many \ac{DRL}-based methods considering joint analog and digital beamforming \cite{Abdelrahman2020TowardsStandaloneOperation, Huang2020DRL_RIS, Feng2020DRL_MISO, Stylianopoulos2022DeepCB}. To this end, we simply set each element of vector $\mathbf{v}$ to $1/K$ to obtain a unit power precoding vector.
\par
To access the quality of the considered communication system, the instantaneous \ac{SNR} performance is defined as (the involved channel matrices need to be perfectly known) $\gamma \triangleq \frac{P}{\sigma^2} \left| \mathbf{g} \myvec{\Phi} \mathbf{H} \right|^2$.
%:
%\begin{equation}
%    \gamma \triangleq \frac{P}{\sigma^2} \left| \mathbf{g} \myvec{\Phi} \mathbf{H} \right|^2.
%\end{equation}
Finally, we formulate the optimization objective of the achievable rate performance per unit bandwidth as a function of the controllable \ac{RIS} configuration and the channel state matrices:
\begin{align}\label{Problem:Max_rate2}
\mathcal{OP}:\quad&\max_{\myvec{\phi}}\;\;R_{\mathbf{g},\mathbf{H}}(\myvec{\phi}) \triangleq \log_2\left(1 + \frac{P}{\sigma^2} \left| \mathbf{g} \myvec{\Phi} \mathbf{H} \right|^2\right)\nonumber   \\
& \quad \hbox{s.t.} \qquad\varphi_i \in \mathcal{F}, \quad  1 \leq i \leq N. \nonumber
\end{align}
In this problem formulation, we make the following assumptions: \textit{i} The channel realizations through time are \ac{IID}; \textit{ii} The \ac{UE} is capable of measuring the received \ac{SNR}, and consequently, share this measure with the RIS controller; \textit{iii} The controller has complete \ac{CSI} knowledge and is capable of changing the configuration of the RIS without delay; and \textit{iv} Similarly, the communication between the controller and the network's nodes is assumed to have negligible effect. The problem of practical channel estimation \cite{RIS_Overview} and its effect on orchestrating RISs lies out of the primary focus of this paper. 
\vspace{-0.1cm}
\section{Proposed DRL-Based RIS Configuration}\label{sec:methodology}
%\vspace{-0.1cm}

\subsection{Reinforcement Learning Formulation} \label{sec:rl-formulation}
The \ac{RL} methodology involves a computationally-enabled agent (i.e., the \ac{RIS} controller) interacting with an environment (i.e., the RIS-enabled communication system) in order to maximize its own goal.
Concretely, at every discrete time step $t$, the agent, upon acquiring an observation from the environment, is tasked with the selection of one of the available actions.
The action affects the internal state of the environment and the latter, in turn, gives off a reward signal to the agent, while moving to the next state.
To formulate an \ac{RL} problem that is equivalent to the design objective in $\mathcal{OP}$ within a finite time horizon, we next define the individual components of the corresponding \ac{MDP}:
\begin{itemize}
    \item \textbf{Observation (state):} We assume that \ac{CSI} is available to the RIS controller, thus the observation vector is consisted of the elements of the two channel matrices:
    $\myvec{s_t} \triangleq [{\rm vec}(\mathbf{H_t}), {\rm vec}(\mathbf{g_t})]^t \in \mathbb{C}^{K(N+1)}$.
    Note that the actual implementations of the \ac{DRL} algorithms are not designed to process complex values, as a result the actual observation vector is constructed as $\myvec{\hat{s}_t} \triangleq [{\rm Real}(\myvec{s_t}^T), {\rm Imag}(\myvec{s_t}^T)]^T$.
    
    \item \textbf{Action:} The agent controls the configuration of the \ac{RIS} by selecting a vector
    $\myvec{a_t}$ with elements in $\{0,1\}^N$,
    so that $[\myvec{a_t}]_i$ sets the phase shift of the $i$-th ($1 \leq i \leq N$) RIS unit element to either of the two predefined phases.
    
    \item \textbf{Reward:} The selected RIS phase configuration is used, which along with the current channel matrix realizations, results in the instantaneous \ac{SNR} measurement at the UE's end. The achievable rate serves as the reward at each discrete time instant $t$:
    $r_t \triangleq R_{\mathbf{g},\mathbf{H}}(\myvec{\phi})$.
    
    \item \textbf{Transition:} At the time $t+1$, the wireless environment proceeds to the next state by sampling \ac{IID} channel realizations as $\mathbf{g}_{t+1}$ and $\mathbf{H}_{t+1}$.
\end{itemize}
The aim of the agent is to maximize the (undiscounted) expected sum of rewards, i.e., $\mathbb{E}\left\{\sum_{t=1}^{T}r_t\right\}$, for some final time horizon $T$. In the previous expression, the expectation is taken with respect to the transition probability distribution.
\par
In the following sections, we proceed under the proposed \ac{MDP} framework to present the modified versions of two \ac{DRL} algorithms, which are named \textit{bin-DQN} and \textit{bin-DDPG}, respectively. For brevity, only the components that are relevant to the proposed modifications are described, omitting implementation details and theoretical understanding to future versions of this work.

\vspace{-0.2cm}
\subsection{bin-DQN}\label{sec:bin-DQN}
%\vspace{-0.1cm}
Q-Learning is a method for solving an \ac{MDP} by finding a (stochastic) policy $\varpi^*$ (i.e., action-selection function) that maximizes the following state-action value function:
\begin{equation}\label{eq:Q}
    Q(\myvec{s},\myvec{a}) \triangleq \mathbb{E}\left\{\sum\limits_{t=1}^{T} r_t \big| \myvec{s_1}=\myvec{s}, \myvec{a_1}=\myvec{a}\right\}.
\end{equation}
The optimal policy results in a Q-function that is maximal for every state-action pair, and for a given state $\myvec{s'}$ it selects the action $\myvec{a'} = \argmax Q(\myvec{s'},\myvec{a})$.
The \acf{DQN} algorithm \cite{DQN} approximates \eqref{eq:Q} by a neural network $Q_{\myvec{w}}(\myvec{s}, \myvec{a})$, with vector $\myvec{w}$ representing its weights, that receives an observation as input and outputs the predicted $Q$ value for each action. The network is trained using \ac{SGD} on the \ac{TD} error function (given the experience samples $(\myvec{s}, \myvec{a}, r, \myvec{s'})$)
\begin{equation}\label{eq:DQN-loss}
    L({\myvec{w}}) = \frac{1}{2}\left(r + \underset{\myvec{\hat{a}}}{\max} Q_{\myvec{w}}(\myvec{s'}, \myvec{\hat{a}}) - Q_{\myvec{w}}(\myvec{s}, \myvec{a}) \right)^2,
\end{equation}
where $\myvec{s'}$ denotes the successor state of $\myvec{s}$ as observed by the agent.
Note that the time and space complexities for selecting an action for the above process is $O(2^N)$, since the network includes one output neuron for each distinct action, i.e., each RIS phase configuration.
\par
\begin{figure}[t]
    \centering
    \includegraphics[clip,trim={1cm 0 0 0},width=0.9\linewidth]{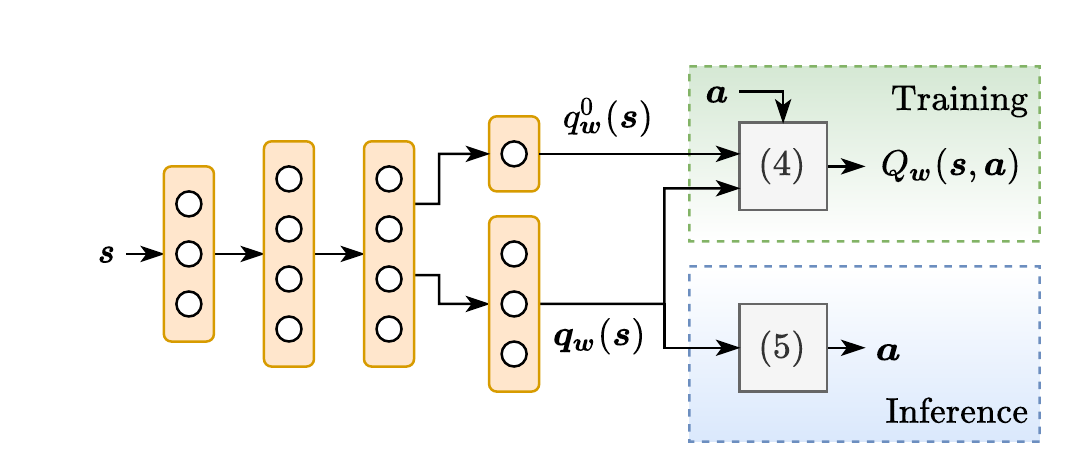}
    \caption{Schematic diagram of the Q-function approximation network of~\cite{dqn-binary-actions}. The action is determined by the use of $[\myvec{q}_{\myvec{w}}(\myvec{s})]_i$ via~\eqref{eq:q-action-selection}, whereas both output layers are utilized to compute $Q_{\myvec{w}}(\myvec{s},\myvec{a})$ through~\eqref{eq:q-approximation} during training updates. This architecture assumes that $\myvec{a}$'s elements take binary values.}\vspace{-0.1cm}
    \label{fig:LAS_DQN_diagram}
\end{figure}

Having defined each action in the previous section as an $N$-element vector with binary elements, in this work, we adopt the neural network architecture proposed in \cite{dqn-binary-actions} that approximates the state-action function as follows:
%\begin{equation}\label{eq:q-approximation}
%    Q_{\myvec{w}}(\myvec{s},\myvec{a}) = q_{\myvec{w}}^0(\myvec{s}) + \sum\limits_{i=1}^N \myvec{a}_{[i]}\myvec{q}_{\myvec{w}}(\myvec{s})_{[i]} = q_{\myvec{w}}^0(\myvec{s}) + \myvec{a}^T \myvec{q}_{\myvec{w}}(\myvec{s}).
%\end{equation}
\begin{equation}\label{eq:q-approximation}
    Q_{\myvec{w}}(\myvec{s},\myvec{a}) = q_{\myvec{w}}^0(\myvec{s}) + \sum\limits_{i=1}^{N} [\myvec{a}]_i [\myvec{q}_{\myvec{w}}(\myvec{s})]_i = q_{\myvec{w}}^0(\myvec{s}) + \myvec{a}^T \myvec{q}_{\myvec{w}}(\myvec{s}).
\end{equation}
The neural network in this expression has two distinct output layers, namely, $q_{\myvec{w}}^0(\myvec{s}) \in \mathbb{R}$ and $\myvec{q}_{\myvec{w}}(\myvec{s}) \in \mathbb{R}^{N\times1}$. It is noted that the dot product operation on the right-hand side of~\eqref{eq:q-approximation} can be seen as a per-element activation/suppression filter on the partial output of the network.
For a given state, the selected action can be derived as:
\begin{equation}\label{eq:q-action-selection}
    [\myvec{a}]_i = \begin{cases}
    1, & [\myvec{q}_{\myvec{w}}(\myvec{s})]_i > 0 \\
    0, & \text{otherwise} \\
    \end{cases}, \quad i=1,2,\dots,N. 
\end{equation}
% For a given state $\myvec{s}$, the selected action can be derived as $\myvec{a}_{[i]} \leftarrow \mathbb{1}_{\myvec{q}_{\myvec{w}}(\myvec{s})_{[i]} > 0}$.
Clearly, the action selected by~\eqref{eq:q-action-selection} is the one maximizing~\eqref{eq:q-approximation} at the given state, since all positive elements of $\myvec{q}_{\myvec{w}}(\myvec{s})$ are activated and contribute to the summation, while the negative values are suppressed.
The architecture is illustrated in Fig~\ref{fig:LAS_DQN_diagram}. Using this approximate form, the space and time complexities of this variation are reduced to $O(N)$, which are tractable to a greater extend, when considering the practical deployment of those algorithms in autonomous wireless systems with potentially limited computational infrastructures.
Let it be noted that~\eqref{eq:q-approximation} remains a differentiable function with respect to $\myvec{w}$, meaning that the gradient steps and backpropagation can be applied as normal under this form.

\vspace{-0.2cm}
\subsection{bin-DDPG}\label{sec:bin-DDPG}
%\vspace{-0.1cm}
While \ac{DQN} has exhibited increased popularity, its domain of application is constrained to problems with discrete action spaces. \acp{MDP}, whose action space is continuous, require special treatment, since it is not straightforward to compute the value function in the continuous domain. Conversely, algorithms of this category are probably better equipped to deal with vector-type actions, since it is typical to employ policy networks that are trained to predict the exact value of each element of the optimal action vector at every time step. As a result, their output value -- when considering the existing \ac{MDP} formulation -- is already shaped as an $N$-element vector. This fact can be exploited to allow for their application in this paper's binary-vector domain, by simply applying a discretization step to turn the continuous values for each element to a binary one.
Concretely, suppose $\myvec{\varpi}_{\myvec{w'}}(\myvec{s}) \in \mathbb{R}^{N\times1}$ denotes the output vector of a policy network when observing state $\myvec{s}$.
Then, the selected action vector can be constructed as:
\begin{equation}\label{eq:discretizationn-action-selection}
    [\myvec{a}]_i = \begin{cases}
    1, & [\myvec{\varpi}_{\myvec{w'}}(\myvec{s})]_i > 0 \\
    0, & \text{otherwise} \\
    \end{cases}, \quad i=1,2,\dots,N,
\end{equation}
which resembles the action selection operation of bin-DQN.
In our numerical evaluation in the next section, we will use the predominant \acf{DDPG} algorithm \cite{DDPG}, upon applying~\eqref{eq:discretizationn-action-selection} to the output of its policy network.
Note that this modification is completely {\em transparent} to the underlying agent and it requires no modifications nor assumptions, apart from the fact that the co-domain of the activation function of the final layer of the policy network contains both positive and negative values, which is readily satisfied by choosing $\tanh$ as the network's activation function.

\vspace{-0.2cm}
\section{Simulation Results}\label{sec:Numerical_Evaluation}
\vspace{-0.1cm}
\subsection{Simulation Setup}
In our simulation setup, we consider the \ac{BS} placed at the 3D Cartesian coordinates $(10,5,2)$, the \ac{UE} at $(8.7, 14.4, 1.6)$, and the \ac{RIS} at $(7.5, 13, 2)$.
The surface is assumed to be oriented parallel to the $y$-$z$ plane.
The predefined RIS phase states per elements are set to $\vartheta_1 = 0$ and $\vartheta_2 = \pi$. The \ac{BS} is operating at the carrier frequency $5~{\rm GHz}$ with transmit power $40~{\rm dBm}$. Additionally, $\sigma^2$ is set to $-100~{\rm dBm}$, and the free space pathloss model is used for computing the attenuation of the channels. The channel coefficients exhibit Ricean fading with a dominant direct path; specifically, the Ricean factors for all involved channels were set to $30~{\rm dB}$. A complete description of the channel model is given in~\cite{Stylianopoulos2022DeepCB}.

\vspace{-0.2cm}
\subsection{Evaluation Process}
\vspace{-0.1cm}
The two modified agents, presented in Section~\ref{sec:methodology}, are tasked to solve $\mathcal{OP}$, by training in the described \ac{MDP}. The two approaches are compared against the original version of the DQN algorithm, along with the random RIS configuration baseline, and the optimal configuration selection strategy, the latter serving as an upper bound. The optimal selection is implemented, when possible, by exhaustively searching all possible configurations for a given channel state.
\par
The agents are trained for $10000$ time steps (i.e., channel realizations).
Since algorithms learn in an off-policy manner (i.e., they purposely select sub-optimal actions at random for better exploration, which provides a lower bound of their true performance), their performance is evaluated at the end of the training process for $1500$ channel transitions, in which the agents act using their (greedy) learned policy.
The baseline and the optimal strategy are evaluated during the same period.
The results, presented below, show the mean rewards (achievable rates) over the evaluation steps.
% The baseline and the optimal strategy are averaged out over $300$ separate realizations at the start of each trial.
Given the plurality of the considered DRL methods and setup instances, we have refrained from performing an extensive hyper-parameter tuning or employing advanced neural network architectures.
The chosen parameter values are given in Table~\ref{tab:hyper-parameters}.
The simulations were performed on a desktop computer with $11$-th generation Intel Core i7 CPU, 32GB RAM, and an Nvidia RTX 3080 (10GB VRAM) GPU. The code was implemented using PyTorch and Tensorflow.
\begin{table}[t]
    \centering
    \caption{Hyper-parameter values of the considered \ac{DRL} algorithms.}
    \vspace*{-5pt}
    %%%%%%%%%%%%%%%%%%%%%%%%%%%%%%%%%%%%%%%%%%%%%%%
    % USE FULL SCREEN TO VIEW THE TABLE CORRECTLY %
    %%%%%%%%%%%%%%%%%%%%%%%%%%%%%%%%%%%%%%%%%%%%%%%
    \begin{tabular}{|l|c|c|c|}
    \hline
    \textbf{Parameter Value}      & \textbf{DQN} & \textbf{bin-DQN}            & \textbf{bin-DDPG}\\
    \hline
    Value network learning rate   &  $0.001$     &  $0.01$                     &  $0.001$         \\
    Policy network learning rate  &  -           &  -                          &  $0.0001$        \\
    Batch size                    &  $128$       &  $128$                      &  $64$            \\
    $\epsilon$-greedy             &  $0.1$       &  $0.1$                      &  -               \\
    Gradient clipping range       & ($-1$, $1$) &  ($-1$, $1$)               &  -               \\
    Target update period          & $1000$       &  $2000$                     &  $1$             \\
    Target soft update temperature& $0.18$       &  $0.05$                     &  $0.00001$       \\
    Ornstein–Uhlenbeck $\mu$      & -            &  -                          &  $0$             \\
    Ornstein–Uhlenbeck $\theta$   & -            & -                           &  $0.15$          \\
    \hline
\textbf{Neural network component} & \textbf{DQN} & \textbf{bin-DQN}            & \textbf{bin-DDPG}\\
    \hline
    Convolutional layers          & \multicolumn{2}{|c|}{$2$}                  & -                 \\
    Units per layer               & \multicolumn{2}{|c|}{$64$}                 & -                 \\
    Kernel width per layer        & \multicolumn{2}{|c|}{$5$}                  & -                 \\
    Max pool layers               & \multicolumn{2}{|c|}{$2$}                  & -                 \\
    Kernel width per layer        & \multicolumn{2}{|c|}{$5$}                  & -                 \\
    Fully connected layers        & \multicolumn{2}{|c|}{$5$}                  & $3$               \\
    Units per layer               & \multicolumn{2}{|c|}{$100$}                & $400$             \\
    Dropout probability           & \multicolumn{2}{|c|}{$0.2$}                & $0.2$                \\
    Activation functions          & \multicolumn{2}{|c|}{$\relu$, $\tanh$}     & $\relu$, $\tanh$  \\
    \hline
    \end{tabular}
    \label{tab:my_label}\vspace{-0.1cm}
    \label{tab:hyper-parameters}
\end{table}

\vspace{-0.2cm}
\subsection{Achievable Rate Evaluation}
\begin{figure}[t]
    \centering
    \includegraphics[width=0.87\linewidth]{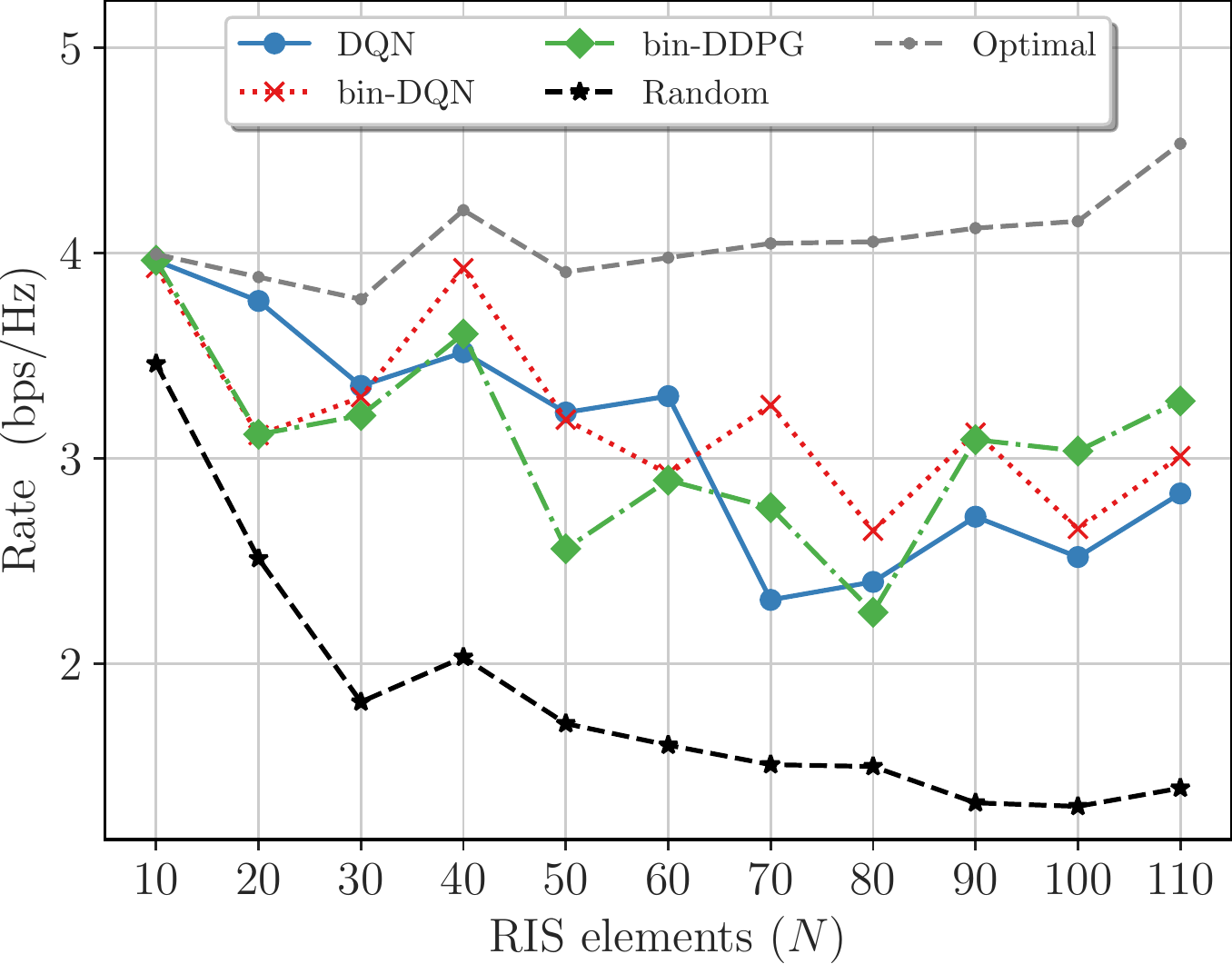}
    \vspace*{-7pt}
    \caption{Achievable rates versus $N$ for the proposed bin-DQN and bin-DDPG in low-to-moderate RIS-sized systems. The original DQN, the (optimal) exhaustive search, and the random configuration are used as baselines.}\vspace{-0.1cm}
    \label{fig:rates-small}
\end{figure}

\begin{figure}[t]
    \centering
    \includegraphics[width=0.91\linewidth]{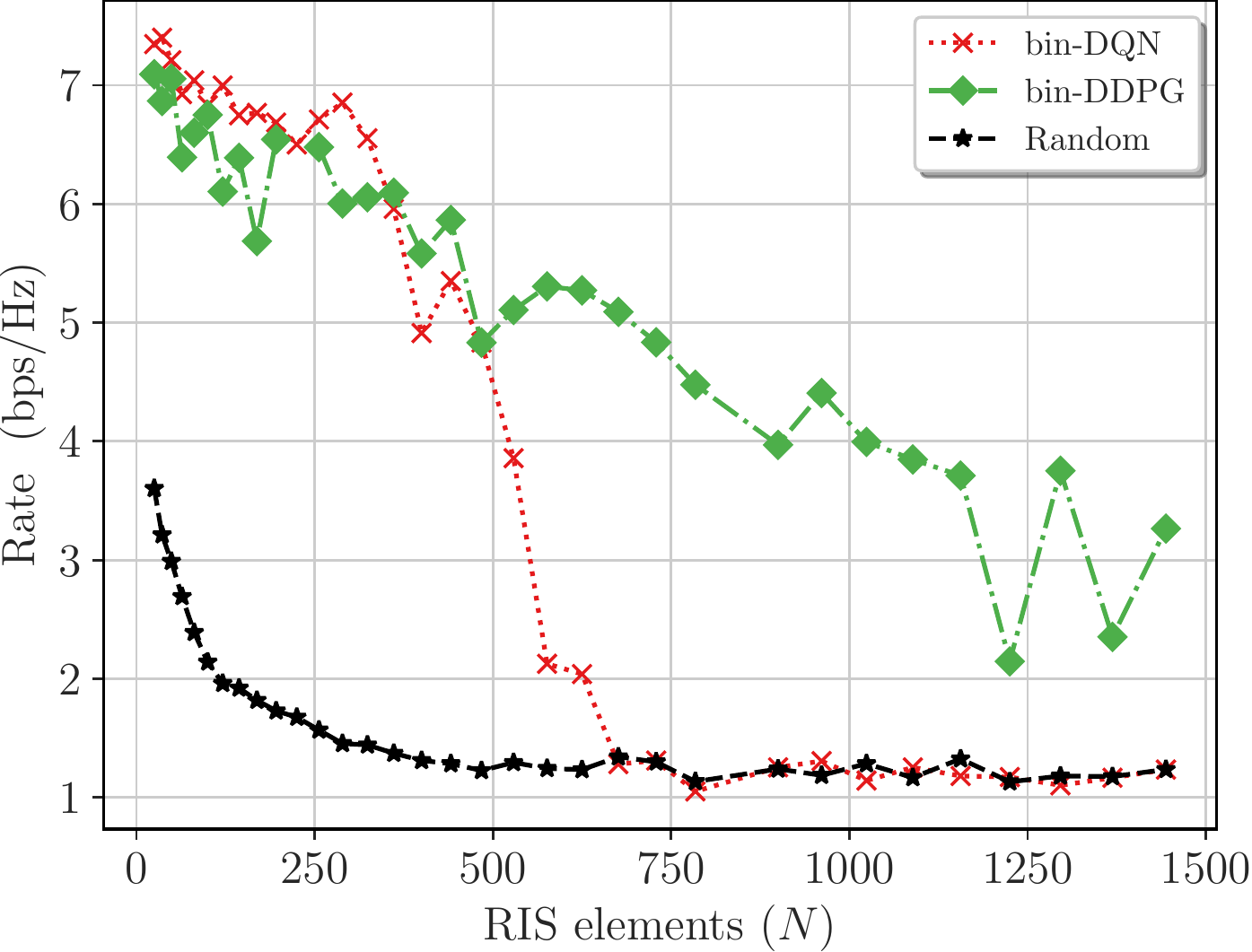}
    \vspace*{-7pt}
    \caption{Achievable rates versus $N$ for the proposed bin-DQN and bin-DDPG algorithms in moderate-to-large RIS-sized systems, compared to the random baseline selection.}\vspace{-0.1cm}
    \label{fig:rates-large}
\end{figure}

\vspace{-0.1cm}
The achievable rate of the proposed DRL algorithms and the benchmark techniques versus the number $N$ of RIS elements are depicted in Figs~ \ref{fig:rates-small} and~\ref{fig:rates-large}.
Due to the exponential increase of both the time complexity of the exhaustive search and the space complexity of the original DQN approach, the computations involved rapidly become intractable.
To account for that, we split the comparison process into two instances. In Fig~\ref{fig:rates-small}, we foremost consider $N$ values up to $110$. In addition, we group $N_{\rm group} = 5$ consecutive RIS elements together, so that elements within the same group share the same reflection phase. Thus, the cardinality of the action space for a given value of $N$ becomes $2^{N/N_{\rm group}}$.
Additionally, note that in order to attain a more extensive collection of evaluation points, we did not constraint the RIS elements to be perfect squares, and as a result, the planar RISs in the simulation have arbitrary rectangular shapes.
Those two modifications affect the behavior of the system under examination, resulting in lower achievable rates with fluctuations in very small RIS sizes. To better investigate the effectiveness of the algorithms in large-scale RISs, we repeat the evaluation process for all perfect square $N$ values up to $1500$ in Fig~\ref{fig:rates-large}. In this figure, only the proposed bin-DQN and bin-DDPG algorithms are displayed, along with the random baseline, since it is infeasible to run exhaustive search or DQN. To account for the enlarged action spaces, we allow the DRL algorithms to be trained for $20000$ time steps; no elements' grouping was applied in that case.
\par
It can be observed from Fig~\ref{fig:rates-small} that the performance of the proposed techniques is on par with the state-of-the-art DQN algorithm for low-to-moderate RIS sizes.
In particular, their performance is close to the optimal achievable rate in the initial toy examples, and at the same time it does not substantially drop, when compared to the naive random RIS configuration selection.% for the largest $N$ values considered in this case.
The results in Fig~\ref{fig:rates-large} showcase that the achievable rates of the randomly configured surface decrease exponentially, which corroborates the need for intelligent configuration techniques in order to benefit from the RIS technology. Interestingly, it is depicted that both proposed DRL agents offer gains in the achievable rate that are double to triple with respect to the random baseline in moderate-to-large RIS sizes (i.e., up to $N=500$). However, the trend in the performance is decreasing, indicating that while effective, the performance of those techniques can be improved. Especially for the case of bin-DQN, it is observed that the drop on its achievable rate is more abrupt, reducing to essentially random action selection for $N>600$. On the other hand, the discretized modification of DDPG continues to exhibit an important improvement over the baseline even for the largest $N$ values, although its behavior is less stable.

% \Kyriakos{
% \textbf{Draft discussion on the results}:
% \textcolor{red}{The general trend in the performance of the algorithms is however decreasing, whereas the optimal rate should increase by a lot at high N values. Does this ruin the whole story? Is it "effective large-scale RIS tuning"?}
% In fact, what this figure shows is that you are better off with a small RIS in the first place, since you cannot effectively control large ones. I suppose this needs addressing.
% }

\vspace{-0.3cm}
\section{Limitations and Future Work}\label{sec:Discussion}
\vspace{-0.1cm}
The performance results in the previous section concern agents incorporating relatively small neural networks in large action and observation spaces (e.g., for $N=1500$, the networks receive $15000$-dimensional state vectors as inputs) and the training periods were restrained to allow for multiple trials to take place.
This is one possible explanation for the general decreasing trend and the fluctuations on DDPG's performance.
At the same time, it is clear that the Q-function approximation adopted here is reasonable up to a certain extend.
A potential extension of this work is to devise a different kind of approximation for the action-state function, that still retains the $O(N)$ complexity, while being tailored to the system modeling details; this direction naturally leads to some variations of deep unfolding \cite{BalatsoukasDeepUnfoldingComms19}.
\par
As discussed earlier, a simplified wireless system has been considered. In practical applications, it is reasonable to assume other free parameters (e.g., precoding selection and power allocation), while having limited \ac{CSI} availability, and intricate network components. Such cases have been covered in the literature often by employing the original versions of the DQN and DDPG algorithms. It is thus our viewpoint that the proposed neural networks can be incorporated as parts of extended purpose agents, with more elaborate \ac{MIMO}-inspired neural network architectures.

\vspace{-0.3cm}
\section{Conclusion}\label{sec:Conclusion}
\vspace{-0.1cm}
Current state-of-the-art DRL approaches suffer from the exponential increase of the action space when large-scale RISs with quantized phases are employed. Our proposed formulation considered $1$-bit resolution phases which allows for the configurations to be viewed as binary-element vectors. Under this viewpoint, we have presented neural network architectures extensions. For the case of DQN, an activation/suppression Q-function approximation was adopted, whereas DDPG's output was discretized. Our simulation results showcased efficient configuration of arbitrary-scale RISs, while providing comparable performance with the considered benchmark approaches.
%The results of the numerical simulations illustrate the the modified versions are effective in controlling large-element surfaces, while providing comparable performance with existing methods in testbed scenarios.

%\section*{Acknowledgements}
%This work has been supported by the EU H2020 RISE-6G project under grant number 101017011.

\FloatBarrier
\bibliographystyle{IEEEtran}
\vspace{-0.1cm}
\bibliography{references}

\end{document}